# Gain sensitivity of the Mach-Zehnder interferometer by photon subtraction strategy

Mikhail S. Podoshvedov[1,2] and Sergey A. Podoshvedov[1*]

[1]*Laboratory of quantum information processing and quantum computing, laboratory of quantum engineering of light, South Ural State University (SUSU), Lenin Av. 76, Chelyabinsk, Russia*
[2]*Kazan Quantum Center, Kazan National Research Technical University named after A.N. Tupolev, Kazan, Russia*
[*]sapodo68@gmail.com

**Abstract:** We study sensitivity of phase estimation of Mach-Zehnder (MZ) interferometer with original two-mode squeezed vacuum (TMSV) state. At the initial stage, the TMSV state is converted into two single-mode squeezed vacuum (SMSV) states, from each of which photons are subtracted by measurement by photon-number resolving (PNR) detector in auxiliary modes. New measurement-induced continuous variable (CV) states of a certain parity can already demonstrate gain sensitivity more than $20\ dB$ in relation to the initial SMSV states at the output from the MZ interferometer and follow to Heisenberg scaling in the case of subtracting a large number of photons in the measuring channels for practical values of the SMSV squeezing $5\ dB >$. Using only one measurement-induced CV state of a certain parity together with the SMSV state shows an increase in sensitivity of no more than $11\ dB$. We show that the sensitivity of the phase estimation obtained by measuring the intensity difference of two measurement-induced CV states in two arms of the MZ interferometer can surpass quantum Cramer-Rao (QCR) boundary of the original two SMSV states just in the practical range of input squeezing $5\ dB >$. In general, the strategy with preliminary subtraction of photons from two SMSV enables greatly enhance the sensitivity of the MZ interferometer in the practical case of small values of squeezing.

## 1. Introduction

Process of measuring a physical system is the main method of understanding its evolution over time. As a result of the measurement, the experimenter can obtain a certain value that can be associated with an assessment of the physical quantity. In general, such an assessment of the measured value is accompanied by a statistical error, which can be both technical (for example, arising due to imperfection of measuring technology) and fundamental (for example, following from Heisenberg's uncertainty principle most often expressed in terms of standard deviations of position and momentum [1,2]. Repeated $m$ measurements provide an average value of unknown parameter with estimation error scaling at best as $\sim m^{-1/2}$ which is a consequence of central limit theorem. The statistical scaling errors is known as standard quantum limit (SQL) [3,4]. If the nonclassical states are used, then the following situation may occur that the uncertainty of the estimate may reach a more fundamental scaling $\sim m^{-1}$, which allows us to speak about the superiority of the obtained estimate by a factor $\sqrt{m}$. Such scaling is known as Heisenberg limit (HL) [5] and allows for one to focus on the number of resources used in the experiment which can be understood as the number of atoms, number of photons, total energy, etc [5]. Feasibility of reaching the Heisenberg limit is intensively discussed [6-8].

Not all nonclassical states can be used to achieve the ultimate limit on estimation precision called Heisenberg limit (HL). One of the most important classes of nonclassical state for ultra-precise estimate of unknown small phase shift are maximally entangled $NOON$ states, the ones with a fixed number of photons [9] whose metrological power can be tested by parity measurement [10]. For the particular case of the $NOON$ states with $N=2$, there exists a deterministic generation recipe by exploiting indistinguishability of photons being quantum resource which interfere on BS through bunching Hong-Ou-Mandel (HOM) effect being purely quantum interference [11]. Violation of SQL with $NOON$ state with $N=2$ via HOM interference has been demonstrated in [12]. HOM-based quantum sensors are used to characterize the single photon sources [13], time delay between two paths [14]. Saturating the Cramer-Rao bound using maximum-likelihood estimator is used to achieve ultra-precise precision in relative arrival time of two photons [15]. Generation of $NOON$ states with $N>2$ is not straightforward, requires a lot of efforts and even the quantum engineering of higher order $NOON$ states may not always guarantee their advantage over other states [16]. Another class of nonclassical states useful for quantum metrology is given by Gaussian states described in CV formalism. An intense research activity is devoted to study their interferometric properties [17,18]. Squeezed states are the most used Gaussian ones to enhance output sensitivity [19-28]. Idea of quantum interferometry to reach sub-SQL sensitivity was proposed in [19] by using squeezed and coherent states as input to the MZ interferometer. It was shown in [21] the intensity difference measurement is not sensitive to any relative phase shift if the TMSV state is fed to the input of the MR interferometer. The estimate of the uncertainty bound of the MZ interferometer with input squeezed-coherent states can be significantly improved so that it reaches the HL [23]. Sub-Heisenberg sensitivity in parity measurement was noted in [26]. A modification of the approach, but with the input Fock states, enables to reach sub-SQL phase uncertainty down to the HL [27]. Squeezed states have been successfully used to improve the sensitivity of LIGO gravitational wave detector [28]. Squeezed light is generated in certain optical nonlinear interactions from original light in coherent state [29]. The squeezed states of light play an important role, at least, in fundamental quantum optics. Despite the routine implementation of the squeezed states of light [30,31] their use is not very widespread due to various difficulties, in particular, some closeness of the squeezed and coherent states in the case of small values of the squeezing amplitude. So, losses of photons may bring a squeezed state of light closer to a coherent one. A proposal to implement correlated squeezed light in a fiber coupler is considered in [32]. Here, we consider the possibility of expanding the capabilities of squeezed states of even with small input squeezing amplitudes in the application to ultra-precise estimation of the unknown phase shift in the MZ interferometer. The approach used is based on measuring quantum systems, resulting in post-measurement state whose initially unitary evolution has been already interrupted by act of knowledge about the quantum system [33], as a means of controlling the final state. There are various interpretations of the mechanism of output state control, for example, as in the protocol of quantum teleportation of an unknown qubit [34], the implementation of which through a hybrid quantum channel is presented in [35]. Enhanced transmission of classical information per optical qubit can be realized by measuring displaced entangled states of light [36]. The measurement of a quantum system can be implemented in an auxiliary mode or modes that can lead to the generation of the desired nonclassical states of light, a technique better known as photon subtraction [37-39]. The extraction of a photon with its preliminary displacement, which can be interpreted as the subtraction of a displaced photon, is considered in [40]. Photon subtraction technique is also applicable to the generation of hybrid entangled states of light [41-45]. Subtraction of photons from either TMSV or SMSV state is a fairly common method of quantum engineering of even/odd optical Schrödinger cat states (SCSs), and if previously the subtraction of one to three photons is considered [46-48], significant advance in the implementation of the output SCS is especially noticeable when extracting a significantly larger number of photons [49]. In general, this approach is quite feasible in practice, taking into account progress in development of photon number resolving (PNR) detectors based on use of transition edge sensors (TESs) in combination with the technology of demultiplication of Fock state with subsequent multiplexing of output signals [50]. Due to the nonunary of the measurement operation the distribution of photons and, as a consequence, nonclassical properties of measurement-induced states can significantly vary. So, protocol of increasing the squeezing in decibels by approximately 1.42 times based on the probabilistic subtraction of two photons and followed by Gaussification procedure is considered in [51]. More accurate squeezing gain follows from [52], where the issue of increasing the sensitivity of measurement-induced CV states of definite parity is also considered. Here, we show the possibility of significant increase more than 20 $dB$ of the sensitivity of the MZ interferometer when using the original TMSV state by extracting a certain number of photons from the initial state. Use of the measurement-induced CV states makes it possible to reach the Heisenberg scaling. The sensitivity of the MZ interferometer pumped by two measurement-induced CV states of a certain parity when measuring the intensity difference can exceed the QCR boundary of the initial two SMSV states which are precisely their origin. It is important to note that the gain sensitivity is precisely achieved in the range of small squeezing amplitudes, that is, in cases that are most often encountered in practice.

## 2. Photon subtraction as a way to increase sensitivity of the MZ interferometer

Due to its simplicity and ease of implementation, the MZ interferometer can already be said to have become a "workhorse" in optical classical metrology. It is a two-beam (two-mode) interferometer consisting of two beam splitters (BSs) [4], one of which splits the input beam of light, and the second mixes the states of light at the exit from the interferometer, thereby realizing a two-channel output light intensity distribution. If two input states are used, then the first interferometry BS also mixes the states. The interferometric channels may differ, which causes an additional phase shift of the corresponding light beam compared to the other. This difference can be initially introduced by inserting a transparent sample capable of transmitting light with less speed but without absorption inside. There are other methods for changing the phase difference between light beams in two optical paths by changing either electric, magnetic fields applied to the material or even the external temperature.

We direct our attention to application of the MZ interferometer evolution pumped by nonclassical CV states to increase the accuracy of estimating the unknown phase shift. Initially the TMSV state is used as in Figure 1. The case of using TMSV state to increase the sensitivity of the MZ interferometer is known [21]. Here, we apply a different method to increase the sensitivity of the MZ interferometer with input TMSV state. The approach can become useful in increasing the sensitivity of estimating the unknown phase when measuring intensity difference at the exit from the MZ interferometer.

In Figure 1, the action begins with the generation of TMSV state with a real squeezing amplitude $r$, i.e. $S(r)_{12}|00\rangle_{12} = exp(r(a_1^+ a_2^+ - a_1 a_2))|00\rangle_{12}$, where the creation $a^+$ and annihilation $a$ operators are also distinguished by mode subscripts 1 and 2, which is deterministically converted by the balanced beam splitter $BS_{12} = \frac{1}{\sqrt{2}}\begin{bmatrix} 1 & -1 \\ 1 & 1 \end{bmatrix}$ into two separable SMSV states with equal modulo but different in sign squeezing amplitudes $s = r$ and $-s$, respectively, as $BS_{12} S(r)_{12} BS_{12}^+ = S(s)_1 S(-s)_2$, where the SMSV operator is given by $S(s) = exp\left(\frac{s}{2}(a^{+2} - a^2)\right)$. Each of the SMSV states passes through additional BS in Fig. 1 with arbitrary real transmission $t$ and reflection $r$ amplitudes

$$BS_{12} = \begin{bmatrix} t & -r \\ r & t \end{bmatrix}, \qquad (1)$$

subject to the normalization condition $t^2 + r^2 = 1$. Additional BSs are introduced to implement quantum engineering of new CV states. Each of the additional BSs transform SMSV state into a complex entangled state represented in equation (A2). In the auxiliary modes of additional BSs, two TES detectors capable of distinguishing from each other the number of incoming photons are placed.

The measurement-induced mechanism realized by accurately measuring the number of photons in the second auxiliary mode of the BS [49,52], makes it possible to generate the following set of CV states of definite parity, i.e. even CV ones

$$|\Psi_{2m}^{(0)}(y_1)\rangle = \frac{1}{\sqrt{Z^{(2m)}(y_1)}} \sum_{n=0}^{\infty} \frac{y_1^n}{\sqrt{(2n)!}} \frac{(2(n+m))!}{(n+m)!} |2n\rangle, \qquad (2)$$

in the case of even measurement outcome and odd CV states

$$|\Psi_{2m+1}^{(0)}(y_1)\rangle = \sqrt{\frac{y_1}{Z^{(2m+1)}(y_1)}} \sum_{n=0}^{\infty} \frac{y_1^n}{\sqrt{(2n+1)!}} \frac{(2(n+m+1))!}{(n+m+1)!} |2n+1\rangle, \qquad (3)$$

with odd outcomes in measurement mode, where the parameters $y_1$ and $B$ are introduced in Appendix A and normalizing factors of the CV states of a certain parity are determined through $2m, 2m+1$ derivatives of the analytical function $Z(y_1) = 1/\sqrt{1-4y_1^2}$, i.e., $Z^{(2m)}(y_1) = dZ^{2m}/dy_1^{2m}$ and $Z^{(2m+1)}(y_1) = dZ^{2m+1}/dy_1^{2m+1}$, respectively. Here in the CV states, the subscript $2m, 2m+1$ is responsible for the number of extracted photons, while the superscript $0$ accounts for for the number of additional input photons. In general, the set (2,3) forms a family of the CV states of a certain parity determined by an additional input state in auxiliary mode of BS (in the vacuum state in the case under consideration) and the states (2,3) themselves are elements of the family.

As can be seen from Figure 1, two SMSV states $|SMSV(y)\rangle_1$ and $|SMSV(-y)\rangle_2$ (Eq. (A1)) simultaneously pass through separate BSs, which in general may differ from each other, but we assume that they are identical. As said earlier, the measuring outputs of the BSs (modes $1'$ and $2'$, respectively) end on PNR detectors each. Simultaneous clicking of both of them in Fig. 1 (say, the measurement outcomes $n_1$ and $n_2$, respectively) guarantees the measurement-induced realization of a separable state consisting of two CV states of a certain parity

$$|\Psi_{n_1,n_2}^{(0,0)}(y_1,-y_1)\rangle_{12} = |\Psi_{n_1}^{(0)}(y_1)\rangle_1 |\Psi_{n_2}^{(0)}(-y_1)\rangle_2. \qquad (4)$$

Note that the difference between the CV states $|\Psi_n^{(0)}(-y_1)\rangle$ and those presented in equations (2,3) is that for them an additional factor $(-1)^n$ is added. The success probability to generate measurement-induced state (4) is determined by their amplitudes as

$$P_{n_1,n_2}^{(0,0)}(y_1,B) = |c_{n_1}^{(0)}(y_1,B)|^2 |c_{n_2}^{(0)}(y_1,B)|^2 Z^{(n_1)}(y_1)Z^{(n_2)}(y_1)/\cosh s. \qquad (5)$$

The state (4) is already the input for the MZ interferometer in Fig. 1. If we do not use additional quantum engineering procedure for the CV states of a certain parity, then, as noted above, the input state to the MZ interferometer is simply separable of two SMSV states

$$|\Psi_{SMSV,SMSV}(s,-s)\rangle_{12} = |SMSV(s)\rangle_1 |SMSV(-s)\rangle_2. \qquad (6)$$

The mathematics of the Mach-Zehnder interferometer is based on the use of quasi-spin operators of a two-mode system [10]

$$J_x = \frac{1}{2}(a_1^+ a_2 + a_1 a_2^+), \quad J_y = \frac{1}{2i}(a_1^+ a_2 - a_1 a_2^+), \quad J_z = \frac{1}{2}(a_1^+ a_1 - a_2^+ a_2), \qquad (7)$$

which obey the commutation relations $[J_i, J_j] = i\epsilon_{ijk} J_k$. When using the MZ interferometer, the BSs are assumed to be balanced, that is, those whose transmission and reflection amplitudes are equal in magnitude to each other. Then the first (input) BS is described by the operator $exp\left(-i\frac{\pi}{2} J_x\right)$ (counterclockwise rotation), while the output BS by the operator $exp\left(i\frac{\pi}{2} J_x\right)$ (clockwise rotation). The phase shift operator of two optical paths has the form $exp(-i\varphi J_z)$, and combination of the three operators accounts for the MZ interferometer output

$$U_{MZ}(\varphi) = exp\left(i\frac{\pi}{2} J_x\right) exp(-i\varphi J_z) exp\left(-i\frac{\pi}{2} J_x\right) = exp(-i\varphi J_y). \qquad (8)$$

Operator $J_y$ is the generator of the unitary transformation (8) used to estimate QCR bound [7,8] for the corresponding input nonclassical states. The quantum Fisher information (QFI) for a pure state is just four times variance of phase evolution generator ($J_y$ in our case). Then, the QFI of the input state (4) takes the form

$$F_{n_1,n_2}(s) = 2(1 + 4y_1^2)\langle n_1\rangle\langle n_2\rangle + (1 + 8y_1^2(n_2 + 1))\langle n_1\rangle + \\ (1 + 8y_1^2(n_1 + 1))\langle n_2\rangle + 8y_1^2(n_1 + 1)(n_2 + 1), \qquad (9)$$

where the symbols $\langle n_i\rangle = \langle n_{n_i}\rangle$ ($i = 1,2$) mean the average number of photons in the CV states (2,3) when measured the number of photons $n_1$ and $n_2$ in auxiliary modes $1'$ and $2'$, respectively. The average number of photons in the measurement-induced $n-$ photon subtracted CV states (2,3) is given by

$$\langle n\rangle = y_1 \frac{Z^{(n+1)}}{Z^{(n)}}. \qquad (10)$$

Here the QFI is defined through the initial squeezing amplitude of the SMSV state, although it is not directly present in the expression (9). In the partial case of the same number $n_1 = n_2 = n$ of photons subtracted, which leads to $\langle n_1\rangle = \langle n_2\rangle = \langle n\rangle$, the quantum Fisher information can be represented in simpler form

$$F_{n,n}(s) = 2\left((1 + 4y_1^2)\langle n\rangle^2 + (1 + 8y_1^2(n + 1))\langle n\rangle + 4y_1^2(n + 1)^2\right), \qquad (11)$$

which obviously shows the superiority of the QFI over the average number of photons squared, that is $F_{n,n}(s) > \langle n\rangle^2$. The QFI of the state (6) becomes equal to

$$F_{SMSV,SMSV}(s) = 2((1 + ctanh^2 s)\langle n_{SMSV}\rangle^2 + \langle n_{SMSV}\rangle) = 4(\langle n_{SMSV}\rangle^2 + \langle n_{SMSV}\rangle), \qquad (12)$$

what may seem even comparable to the expressions (9,11).

Important aspect of optical interferometry is its phase sensitivity that is quantified by an average value of how much the measured phase could differ from the actual value $\Delta\varphi$. The most sensitive measurement has the smallest $\Delta\varphi$. It is well known the smallest phase uncertainty could not be smaller of QCR bound determined through the QFI [1]. As the CV states under study (4,6) can carry distinct QFI it can result in different maximum sensitivities of the MZ interferometer. To compare them, in Figs. 2(a-d) and Figs. 3(a-d), we present dependencies of the QCR bound

$$\Delta\varphi_{n_1,n_2} = \frac{1}{\sqrt{F_{n_1,n_2}(s)}}, \quad \Delta\varphi_{n,n} = \frac{1}{\sqrt{F_{n,n}(s)}}, \quad \Delta\varphi_{SMSV,SMSV} = \frac{1}{\sqrt{F_{SMSV,SMSV}(s)}}, \qquad (13)$$

on the squeezing amplitude $S$ ($dB$) of the original TMSV state. It follows from definition of the QFI of presented CV states, the QCR bound of both photon subtracted and original SMSV states is independent on the unknown estimated phase shift $\varphi$. Plots in Figs. 2(a-d) are constructed for different values of $n_1$ and $n_2$ and the same transmission amplitude $t = 0.9$ of the BSs used in quantum engineering of new CV states of definite parity. In Figure 3(a-d), QCR bounds are shown already in the case of the same number $n_1 = n_2 = n$ of photons subtracted (total number of subtracted photons in both channels is $2n$) but for different transmission amplitudes $t$. The condition $\Delta\varphi_{SMSV,SMSV} > \Delta\varphi_{n_1,n_2}$ or $\Delta\varphi_{SMSV,SMSV} > \Delta\varphi_{n,n}$ is fulfilled in a wide range of changes $S$ starting with $S_1 = 0$ up to certain big enough value $S_2$ which is evidence of the usefulness of preliminary subtraction of photons from two SMSV states in the diapason of $S$. The value of $S_2$ strong enough depends on the number of photons subtracted and to a lesser extent on the transmission amplitude $t$. For the values of $S \geq S_2$, the condition either $\Delta\varphi_{n_1,n_2} \geq \Delta\varphi_{SMSV,SMSV}$ or $\Delta\varphi_{n,n} \geq \Delta\varphi_{SMSV,SMSV}$ is observed, which may already indicate the uselessness of the photon subtraction procedure from the SMSV states in the light of an increase of the sensitivity of the MZ interferometer. What is important to note here is that the detected sensitivity gain range covers practical values of $S$, that is, those that can be implemented in a laboratory case. So in the most common practical case of $S < 5\ dB$ [25], an even stronger reduction of the noise can be demonstrated in the MZ interferometer, that is, the condition $\Delta\varphi_{SMSV,SMSV} \gg \Delta\varphi_{n_1,n_2}$ and $\Delta\varphi_{SMSV,SMSV} \gg \Delta\varphi_{n,n}$ performance is visually observed, which is more evident in favor of the use of photonic subtraction in the practical case of small values of the squeezing amplitude $S$.

In Figures 2 and 3 we also present functions comparable to the Heisenberg limit as well as SQL, namely, $1/(\langle n_1\rangle + \langle n_2\rangle)$ and $1/\sqrt{\langle n_1\rangle + \langle n_2\rangle}$ in Figs. 2 and $1/(2\langle n\rangle)$ and $1/\sqrt{2\langle n\rangle}$ in Figs. 3 to judge the scaling of the scheme with increasing number of particles. Here the addition of the average number of photons occurs due to due to the fact that two CV states are used at the entrance to the MZ interferometer. The curves are not presented for all values $n_1, n_2$ and $n$ used when constructing the QCR boundary, but in general presented dependencies are enough when comparing. From the data obtained, it is noticeable that the QCR boundary of CV states of definite parity is significantly less than QSL and can significantly approach the Heisenberg limit, especially in the case of using $n, n -$ photon subtracted CV states, when the curves match with high accuracy with $n$ growing. In general, one can estimate the QCR bound of photon-subtracted CV states as being in the range either $1/(\sqrt{\langle n_1\rangle + \langle n_2\rangle}) > \Delta\varphi_{n_1,n_2} \geq 1/(\langle n_1\rangle + \langle n_2\rangle)$ or $1/\sqrt{2\langle n\rangle} > \Delta\varphi_{n,n} \geq 1/(2\langle n\rangle)$. Note that the last expression in the equation (12) obviously talks about sub-Heisenberg sensitivity $1/2\langle n_{SMSV}\rangle >$ of the MZ interferometer which is pumped by two SMSV states (6).

In general, three main features appear in the graphical dependencies. Firstly, an increase in the value of the parameter $S$ leads to a decrease in the values of the QCR limit no matter which CV states of certain parity are used at the entrance to the MZ interferometer. Secondly, increasing the number of photons subtracted also monotonically decreases the limiting uncertainty $\Delta\varphi$ in estimating the unknown parameter. Since in the case of plotting the curves in Figure 3, the total number of photons subtracted doubles ($2n$ instead of $n_1 + n_2$), therefore, in a rough approximation, better sensitivity is achieved. The graphic dependences contain a third factor associated with the transmission amplitude $t$ of the BSs used to create measurement-induced CV states. Increasing the parameter $t$ can also increase the sensitivity of the MZ interferometer although to somewhat lesser extent. But it comes at the cost of reducing the success probability (5) in generating the CV states of definite parity. Finally, an increase in both the number of photons subtracted from the initial SMSV states and the transmission amplitude $t$ of the BS guarantees the movement of the parameter $S_2$ towards larger values along the horizontal axis, thereby increasing the sensitivity of the MZ interferometer. As can be seen from the graphs in Figs. 2 and 3, the parameter $S_2$ may achieve values of $20 - 30\ dB$. The generation of the squeezed light with squeezing amplitude greater than $> 20\ dB$ is known to present significant technical difficulties and is hardly possible at the present time. Therefore, the strategy of initial subtraction of a certain number of photons from the SMSV states with small squeezing is more attractive as allows for the MZ interferometry to achieve estimation accuracy unattainable with two original SMSV states with much greater squeezing.

To save experimental resources, consider the quantum uncertainty in the case when photons are subtracted only in one channel (assume in first in Fig. 1) while the state in the adjacent channel remains the same, that is, the SMSV state in Fig. 1

$$\left|\Psi^{(0)}_{n_1,SMSV}(y_1,-s)\right\rangle_{12} = \left|\Psi^{(0)}_{n_1}(y_1)\right\rangle_1 |SMSV(-s)\rangle_2. \tag{14}$$

Then, the QCR bound of the MZ interferometer

$$\Delta\varphi_{n_1,SMSV} = \frac{1}{\sqrt{F_{n_1,SMSV}(s)}}, \tag{15}$$

is determined through the quantum Fisher information

$$F_{n_1,SMSV}(s) = 2(1 + 2y_1 ctanhs)\langle n_1\rangle\langle n_{SMSV}\rangle + \langle n_1\rangle + (1 + 4y_1 ctanhs(n_1 + 1))\langle n_{SMSV}\rangle. \tag{16}$$

The corresponding dependences of the maximum sensitivity of the MZ interferometer with the input state (14) as function of the input squeezing $S$ $(dB)$ are shown in Figure 4(a-d) for different values of $n_1$ and $t$. Comparison of the dependencies with those stemming from the case with CV state (4) shows deterioration of the phase sensitivity when photon subtraction occurs only in one channel especially at the initial stage of small squeezing under other identical conditions. An explanation may be given by a decrease in the total number of photons subtracted. Here, as in Figures 2 and 3, there is the phenomenon of intersection of the curve $n_1 - SMSV$ with other curves $n_1 - n_2$ at point $S_2$. The point $S_2$ at which the curves intersect with the $SMSV - SMSV$ curve can also depend significantly on the number $n_1$ of photons subtracted and the transmission amplitude $t$ and, in general, the point $S_2$ is shifted towards larger values, that is, least $S_2 > 20\ dB$. As for the Heisenberg limit, it lies below the $n_1 - SMSV$ curves.

To compare quantitatively sensitivity gain of $n_1, n_2 -$ subtracted CV state (4) with respect to $n_1 -$subtracted together with SMSV states (14) in Figs. 5(a-d) we show dependencies of $g_{n_1,n_2/n_1,SMSV} = -10\ log(\Delta\varphi_{n_1,n_2}/\Delta\varphi_{n_1,SMSV})\ (dB)$ (or just $g_n$ as indicated in Figure 5) on input squeezing amplitude $S$ $(dB)$ for different values of $n_1$ and $n_2$. The gain sensitivity can be quite complex function from the parameters $n_1, n_2, S$ and $t$. The gain sensitivity starts from $g_n = 0$ at the point $S = 0$, reaching the maximum possible value $g_{max}$ in the vicinity of practical input squeezing $S \approx 5\ dB$, after which it drops to zero and then decreases towards negative values indicating that the sensitivity of the state (14) already exceeds the one of the state (4). It means that the photon subtraction strategy to increase the sensitivity of the MZ interferometer no longer works if $S > S_2$. Maximal gain sensitivity of $n_1, n_2 -$ photon subtracted CV state relative to $n_1 -$ photon CV state around $g_{max} \approx 11\ dB$ can be achieved. It is interesting to note that the observed maximum sensitivity of $n_1, n_2 -$ photon subtracted CV state with respect to the initial two-mode separable SMSV state (6) already reaches values of more than $> 20\ dB$ which allows us to talk about an approximately twofold increase in sensitivity when using two photon subtracted CV states instead of one $n -$ photon subtracted CV state. The dependence of the gain sensitivity on the number $n_1$ of subtracted photons is observed in Figure 5, but the growth already decreases with an increase in extracted photons. Therefore, from practical point of view, a strategy with small number of photons subtraction is second channel (say, 3 photons) in addition to some number (say, 8 photons) of subtracted photons in the first channel can be used to reach gain sensitivity $\sim 10\ dB$ for $S = 5\ dB$. With larger number of the subtracted photons (say, $n_1 \leq 30$ photons) gain sensitivity is in the region of $\sim 15\ dB$, which corresponds to a $31.65 -$fold increase of sensitivity of phase estimation with the same input squeezing $S = 5\ dB$ achievable in practice.

Thus, the strategy with preliminary subtraction of photons from original SMSV states stemming from TMSV state to enhance the sensitivity of the MZ interferometer becomes actual. Enhancement of the phase sensitivity of the MS interferometer can be achieved in a number of ways, including subtracting photons from both one and two channels, by changing the transmittance of the beam splitters used for the photon subtraction as well as varying input squeezing. QFI $F_{n_1,n_2}(s)$ in Eq. (9) can reach the maximum possible value, in particular, due to additional term proportional $(n_1 + 1)(n_2 + 1)$ which can take on fairly large values, especially, if $n_1$ and $n_2$ increase. Therefore, the method with $n_1, n_2 -$ subtracted photons CV states can be used to obtain the highest possible sensitivity of the phase estimation in MZ interferometer and the phase uncertainty can reach the Heisenberg limit.

## 3. Phase estimation when measuring intensity difference by measurement-induced CV states

Traditional interferometric setup typically measures the intensity difference $D = n_1 - n_2$ at the output corresponding to the visibility of the interference fringes. The intensity difference measurement, which is standard for optical interferometry with coherent state, must phase sensitive with the input states [19,23]. The unknown phase shift between the arms of the MZ interferometer in Fig. 1 is estimated from the intensity difference at output ports and sensitivity of such measurement is quantified by mean value $\Delta\varphi$ of how much the estimated phase could differ from the actual value. Since sensitivity of the $n-$ photon subtracted states can be quite high, it is interesting to test the metrological capabilities of the MZ interferometry when measuring the output intensity difference with the input state (4).

The evolution of the operator $J_z$ in the Heisenberg representation is shown in Eq. (B1). If the same number of photons $n_1 = n_2$ are subtracted from the initial SMSV states, then the output average difference measurement remains zero. In particular, the measured intensity difference of two SMSV states with equal modulo squeezing amplitudes is not sensitive to the phase shift. The variance of the output operator $J_z$ is presented in Eqs. (B3-B5). Error propagation formula gives a good approximation of the sensitivity of the phase estimation

$$\Delta\varphi_\rho = \frac{\Delta D_\rho}{\left|\frac{\partial \langle D \rangle_\rho}{\partial \varphi}\right|}, \tag{17}$$

where double value of the average $\langle D \rangle_\rho = 2\langle J_z \rangle$ and its standard deviation $\Delta D_\rho = 2\Delta J_z$ of the operator $J_z$ is used and subscript $\rho$ points to the state for which the statistical characteristics of the operator $D$ are found.

Output average intensity difference $\langle D \rangle_\rho$ depends as $cos\varphi$ on the phase shift as $\langle J_X \rangle = 0$ and estimation (17) predicts large phase fluctuations around the points $\varphi = 0, \pi$. Optimal point for better phase estimation can follow from the phase shift $\varphi = \pi/2$. Indeed, numerical estimates of $\Delta\varphi$ in dependency on the phase shift $\varphi$ in Figure 6(a-d) confirm this pointing to $\varphi = \pi/2$ as the value at which it is possible to estimate the phase difference with a smallest uncertainty $\Delta\varphi$. The error propagation formula gives a rather complex dependence of the phase estimation on the number of subtracted photons (moreover, there is a dependence on the number of the subtracted photons in both the first $n_1$ and second $n_2$ measuring channels), the squeezing amplitude $S$ of the TMSV state, as well as the transmission amplitude $t$ of the BSs used in implementing the photon subtraction. A fairly small photon subtraction (say, 4 photons in first channel) does not significantly reduce the estimation error at point $\varphi = \pi/2$. The number of subtracted photons in the first channel is required to increase to significantly improve the sensitivity of the phase estimation. In general, other parameters being equal, increasing the number of subtracted photons (say up to $n_1 = 10$ photons in first measurement port) can significantly increase the sensitivity of MZ interferometry when estimating phase shift $\varphi = \pi/2$ by measuring output intensity difference. However, note that increasing the number of subtracted photons (say to 6 photons) in the second channel may reduce the increase in sensitivity (Fig. 6(d)) relative to the cases with fewer subtracted photons in that channel.

The sensitivity of the phase estimation of the phase shift $\varphi = \pi/2$ when measuring intensity difference at output of MZ interferometer output as a function of the squeezing amplitude $S$ ($dB$) is presented in Figs. 7(a-d) under other used parameters being equal. Although a rather complex dependence of formula (17) on $n_1, n_2, S$ and $t$ one can definitely talk about an increase in the sensitivity of the phase estimation obtained in the MZ interferometer with increasing the initial squeezing $S$. Traditionally, sensitivity significantly increases with the number of subtracted photons, e.g. $n_1, n_2$ growing. A small number of subtracted photons (say, $n_1 = 2$) may not guarantee an increase in the sensitivity when measuring the intensity difference at the exit from the MZ interferometer. In addition, Figure 7(a-d) also shows the dependence of the QCR boundary of two SMSV states (6) that were previously shown in Figures 2-4. As can be seen, the sensitivity of phase estimation when intensity difference of measurement-induced CV states is measured surpass QCR bound of two SMSV states, although $\Delta\varphi_\rho > \Delta\varphi_{n_1,n_2}$. At least the superiority of sensitivity occurs precisely in the region of practical squeezing values $< 10\ dB$. Only starting from a certain value of $S_{SMSV}$, the QCR boundary of two SMSV states becomes less than the one achieved by $n_1, n_2 -$ photon subtracted CV states. Overall, the value of $S_{SMSV}$ depends on $n_1$ and $n_2$ and therefore allows for manipulation of the MZ interferometer sensitivity by subtracting a corresponding number of photons in both measurement channels in Fig. 1.

## 4. Conclusion

Quantum engineering of new CV states by subtracting photons from the initial state is attractive from the point of view of it practical implementation [37-39]. As a result, the distribution of photons in the state subject to non-unitary operation can significantly change which can greatly change nonclassical properties of the photon-subtracted state. However, removing a photon from a coherent state does not change the input state. If photon subtraction applies to thermal state, it may seem surprising but the average number of photons in a photon-subtracted thermal state only doubles. The quantum engineering of the CV states based on photon subtraction [46-48] is not limited to the examples taking into account the successful development of photon-number resolving technology which enables to distinguish from each other numbers of photons significantly larger than one photon [50]. Here we use measurement-induced CV states of certain parity for the purposes of optical quantum metrology. We provided a brief discussion of conversion of original TMSV state into input state of two CV states of definite parity which is used as input to the MZ interferometer. In general, three types of separable CV states, namely two CV states of a certain parity, one CV state of a certain parity together with SMSV state and both SMSV states are used to evaluate their potential usefulness in ultra-precise estimations of the phase shift. After that, we applied the basic formalism of quantum metrology based on the QCR bound to the states. For all quantum states considered, sub-SQL phase sensitivity is achieved, while a pair of the CV states of a certain parity can reach the Heisenberg limit in the case of large number of subtracted photons and two SMSV states can even overcome the Heisenberg scaling. Despite the fact that the implementation of both TMSV and SMSV states is currently a fairly routine task in world laboratories, the issue of increasing its squeezing properties, which is directly related to the squeezing parameter $S$, is an intractable problem. This greatly limits the use of the SMSV states in optical quantum metrology due to its small average number of photons. Quantum engineering based on subtraction of a certain number of photons from SMSV states enables to overcome their disadvantage and increase the sensitivity of the phase estimation by more than $20\ dB$ just in the region of small values of squeezing up to $5\ dB$ of original nonclassical CV states. Even extraction of the small number of photons, not to mention a subtraction of significant number of photons, can guarantee an acceptable sensitivity gain. Using only one measurement-induced CV state of a certain parity, leaving the SMSV state in the adjacent channel, reduces the sensitivity of the MZ interferometer by approximately two times. As follows from the analysis, the sensitivity of the phase estimation is determined not only by the number of extracted photons beyond the initial squeezing, but also by the transmission amplitude of BSs with the help of which the measurement-induced states are generated.

  Measuring the intensity difference at the output of the MS interferometer can be a standard technique to test measurement-induced CV states for the phase estimation. The measurement is not saturating, but it is worth noting that an increase in the number of extracted photons even enables to overcome the QCR boundary of two original SMSV states from which they are realized just with a small initial squeezing $5\ dB >$. In a practical implementation, a set of statistical results is collected for different values of the subtracted photons, which allows for the uncertainty of the phase difference to be estimated for all observed measurement events. Thus, the photon subtraction approach can improve the metrological power of the MS interferometer at a level superior QCR bound of two SMSV states (or TMSV state) with realistic squeezing amplitudes and the technique can become practically feasible taking into account progress in the development of TES technology. In principle, one may measure corresponding observable $J_y$ for the state (4) to gain definite access to phase differences $\Delta\varphi$ which deserves special consideration.

## Appendix A. Passage of the SMSV state through a beam splitter

Here we show the output state from the BS in equation (1) if the SMSV state

$$|SMSV(y)\rangle = \frac{1}{\sqrt{\cosh s}}\sum_{n=0}^{\infty}\frac{y^n}{\sqrt{(2n)!}}\frac{(2n)!}{n!}|2n\rangle, \quad (A1)$$

is launched to one of its inputs, leaving another empty. Here, an amount $y = \tanh s/2$ is a squeezing parameter, $s > 0$ is the squeezing amplitude which provides the range of its change $0 \leq y \leq 0.5$. The value of the squeezing parameter $y = 0$ indicates the absence of the SMSV state at the input to the BS, while the value $y = 0.5$ corresponds to the physically unrealizable case of a maximally squeezed vacuum state with amplitude $s \to \infty$. In addition to the two SMSV parameters, two other parameters, namely, the squeezing expressed in decibels $S = -10\log(\exp(-2s))\, dB$ and the average number of photons $\langle n_{SMSV}\rangle = \sinh^2 s$ can also be used when describing the SMSV state used. The BS mixes modes 1 and 2 transforming the creation operators $a_1^+$ and $a_2^+$ as $BS_{12}a_1^+BS_{12}^+ = ta_1^+ - ra_2^+$ and $BS_{12}a_2^+BS_{12}^+ = ra_1^+ + ta_2^+$, respectively, It converts the original SMSV state into a hybrid entangled state [49,52],

$$BS_{12}(|SMSV(y)\rangle_1|0\rangle_2) = \frac{1}{\sqrt{\cosh s}}\sum_{n=0}^{\infty}c_n^{(0)}(y_1,B)\sqrt{Z^{(n)}(y_1)}|\Psi_n^{(0)}(y_1)\rangle_1|n\rangle_2, \quad (A2)$$

with amplitudes

$$c_n^{(0)}(y_1,B) = (-1)^l\frac{(y_1B)^{\frac{n}{2}}}{\sqrt{n!}}, \quad (A3)$$

where the input squeezing parameter $y$ decreases by $t^2$ times, that is, it becomes equal to $y_1 = yt^2 = y/(1+B) \leq y$ and the BS parameter is equal to $B = (1-t^2)/t^2$. CV states of a certain parity as well as the designations used are presented in the main part of the manuscript.

If two separate BSs act separately on one of the two SMSV states as shown in Fig. 1 then the final state becomes

$$BS_{11'}(|SMSV(y)\rangle_1|0\rangle_{1'})BS_{22'}(|SMSV(-y)\rangle_2|0\rangle_{2'}) = \frac{1}{\cosh s}$$

$$\sum_{n_1=0}^{\infty}\sum_{n_2=0}^{\infty}\begin{pmatrix}c_{n_1}^{(0)}(y_1,B)c_{n_2}^{(0)}(-y_1,B)\sqrt{Z^{(n_1)}(y_1)Z^{(n_2)}(y_1)}\\ |\Psi_{n_1}^{(0)}(y_1)\rangle_1|\Psi_{n_2}^{(0)}(-y_1)\rangle_2\end{pmatrix}|n_1\rangle_{1'}|n_1\rangle_{2'}, \quad (A4)$$

where the amplitude $c_{n_2}^{(0)}(-y_1,B)$ is equal to $(-1)^{m_2}c_{2m_2}^{(0)}(y_1,B)$ if $n_2 = 2m_2$ is even and $(-1)^{m_2+1}c_{2m_2+1}^{(0)}(y_1,B)$ if $n_2 = 2m_2 + 1$ is odd.

## Appendix B. Output characteristics of the operator $J_z$

Schwinger realization of the $SU(2)$ Lie algebra is employed to calculate detection observables [10]. Based on definitions of two-mode operators, one can obtain input-output relations between $J_{zIn}$ and $J_{zOut}$

$$J_{zOut} = U_{MZ}^+(\varphi)J_{zIn}U_{MZ}(\varphi) = \exp(i\varphi J_y)J_{zIn}\exp(i\varphi J_y) = \cos\varphi J_{zIn} - \sin\varphi J_{xIn}, \quad (B1)$$

which provides

$$\langle J_{zOut}\rangle = \cos\varphi\langle J_{zIn}\rangle, \quad (B2)$$

due to operator $J_x$ parity failure, i.e. $\langle J_x \rangle = 0$, where $\langle J_{zIn} \rangle = \frac{1}{2}(\langle n_1 \rangle - \langle n_2 \rangle)$. Again using the parity of the CV states (2,3) used, one obtains the output dispersion of the operator $J_z$ for the states

$$(\Delta J_{zOut})^2 = \frac{1}{4}(cos^2\varphi((\Delta n_1)^2 + (\Delta n_2)^2) + 4sin^2\varphi(\Delta J_x)^2), \qquad (B3)$$

where $(\Delta n_1)^2 = \langle n_1^2 \rangle - \langle n_1 \rangle^2$ and $(\Delta n_2)^2 = \langle n_2^2 \rangle - \langle n_2 \rangle^2$ are the dispersions of the number of photons in $n-$photon subtracted CV state

$$(\Delta n_i)^2 = y_1^2 \frac{Z^{(n+2)}}{Z^{(n)}} + \langle n_i \rangle - \langle n_i \rangle^2 \qquad (B4)$$

with $n = 1,2$, while dispersion of the operator $J_x$ for the input state (4) becomes

$$(\Delta J_x)^2 = \frac{1}{4}\begin{pmatrix} 2(1 - 4y_1^2)\langle n_1 \rangle\langle n_2 \rangle + (1 - 8y_1^2(n_2 + 1))\langle n_1 \rangle + \\ (1 - 8y_1^2(n_1 + 1))\langle n_2 \rangle - 8y_1^2(n_1 + 1)(n_2 + 1) \end{pmatrix}, \qquad (B5)$$

which is converted into

$$(\Delta J_x)^2 = \frac{1}{2}\left((1 - 4y_1^2)\langle n \rangle^2 + (1 - 8y_1^2(n + 1))\langle n \rangle - 4y_1^2(n + 1)^2\right), \qquad (B6)$$

in the case of $n_1 = n_2 = n$.

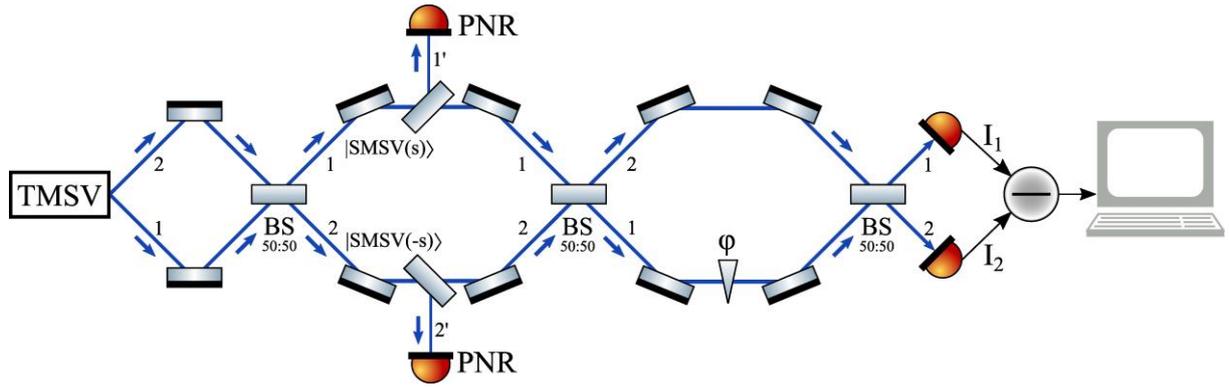

**Fig. 1.** Scheme used to enhance the phase estimation sensitivity obtained with intensity difference measurement $(I_1 - I_2)$ at the exit from the MZ interferometer. Initially, the TMSV state is launched to the balanced beam splitter for conversion into a separable state from two SMSV states. Two PNR detectors are used to extract a certain number of photons (say, $n_1, n_2$) in the corresponding channels, after which the resulting measurement-induced state is fed to the input of the MZ interferometer. The strategy of subtracting photons from the initial SMSV states justifies itself because enables to significantly enhance phase sensitivity of the MZ interferometer to more than $20\ dB$.

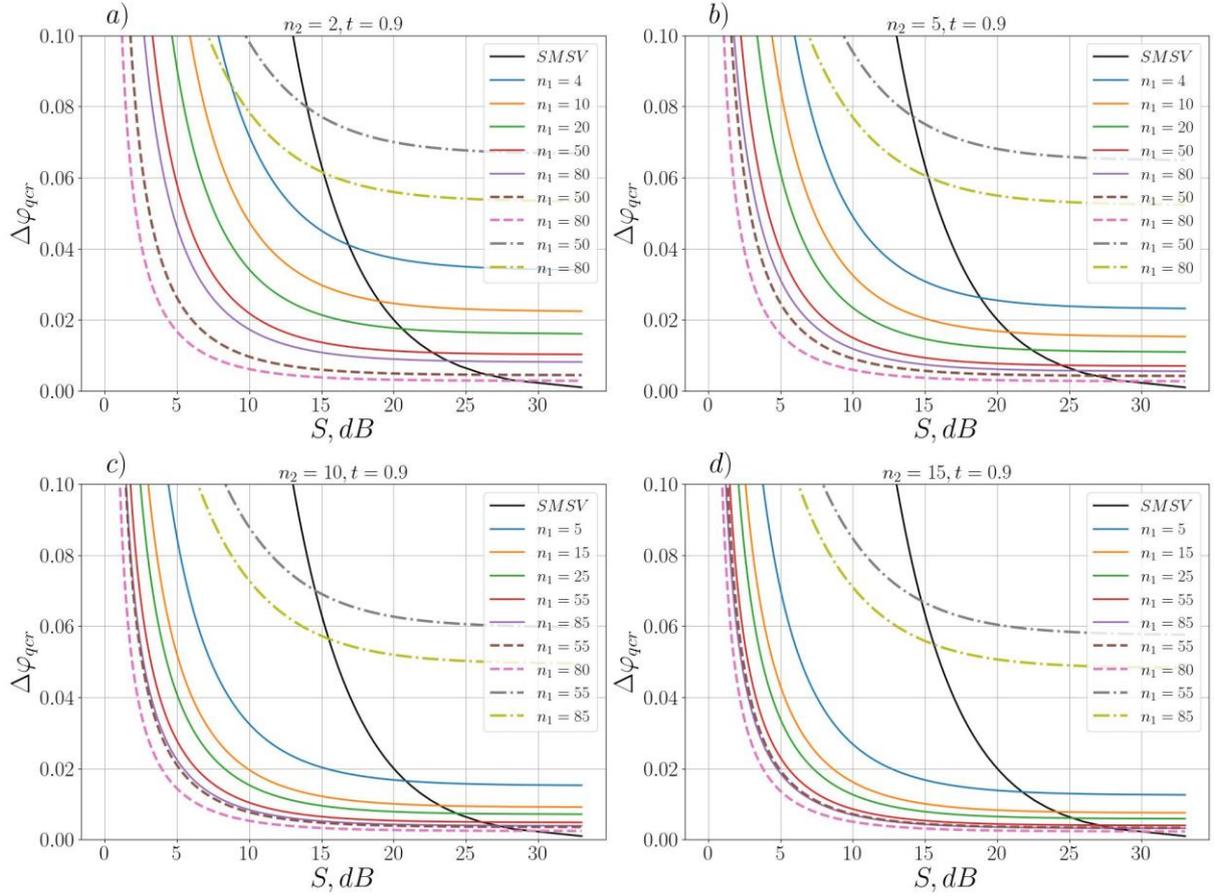

**Fig. 2(a-d).** Dependence of the QCR boundary $\Delta\varphi_{n_1,n_2}$ (here and in other figures abbreviation $\Delta\varphi_{qcr}$ along the vertical axis with subscript $qcr$ applies to all cases in contrast to the designations used in the text) when estimating the unknown phase shift on the squeezing amplitude $S$ ($dB$) of the initial TMSV state in Fig. 1 in the case when two even CV states (Eq. (4)) are simultaneously fed to the input of the MZ interferometer for different values of $n_1$ and $n_2$ (total number of subtracted photons in both used channels is $n_1 + n_2$). The case $\Delta\varphi_{SMSV,SMSV}$ when two SMSV states (6) are simultaneously mixed in the interferometer is also presented (graph in black with designation $SMSV$). Different values of $t$ are used. The dashed lines show the Heisenberg limit and dash-dot lines are the SQL curves for some partial numbers $n_1$ and $n_2$.

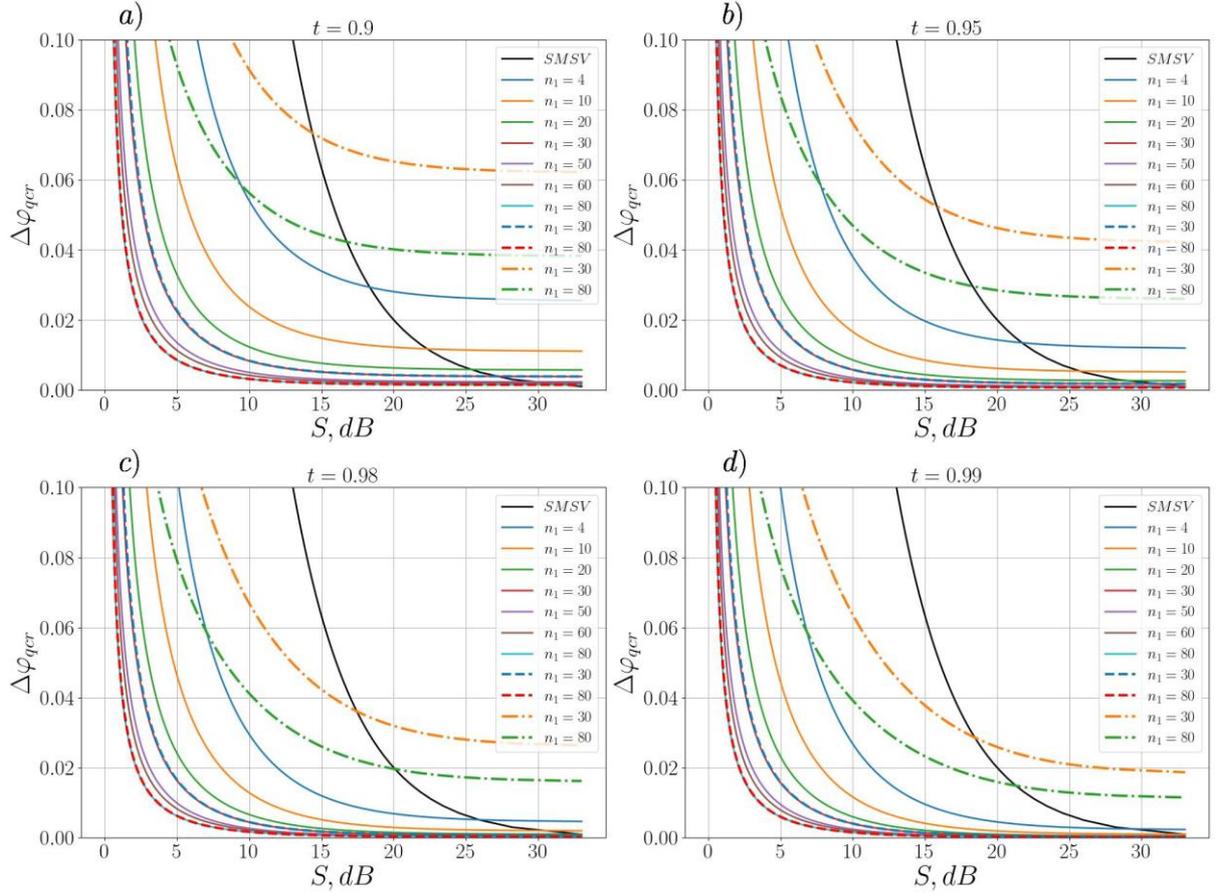

**Fig. 3(a-d).** Maximal sensitivity (QCR boundary) obtained in MZ interferometer $\Delta\varphi_{n,n}$ together with $\Delta\varphi_{SMSV,SMSV}$ as a function of the squeezing amplitude $S$ ($dB$) of the original TMSV states for different values of $n_1 = n_2 = n$ (total number of subtracted photons in both used channels is $2n$). Overall, enhancement in sensitivity compared to those presented in Figure 2 is observed since a larger number of photons are subtracted from SMSV states. The sensitivity of the $n-$ photon subtracted CV states reaches the Heisenberg scaling (dashed lines) and significantly surpass SQL (dash-dot lines).

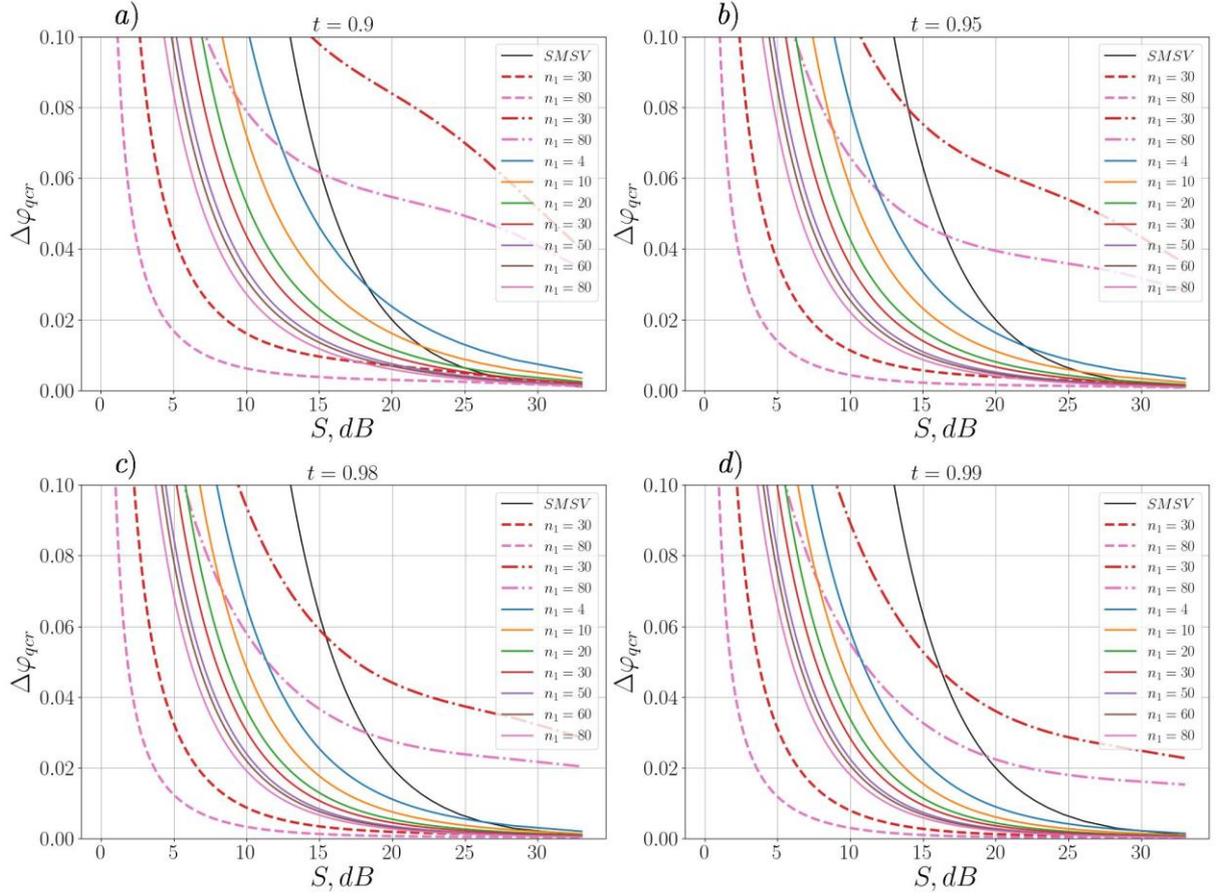

**Fig. 4(a-d).** Dependence of the QCR boundary $\Delta\varphi_{n_1,SMSV}$ and $\Delta\varphi_{SMSV,SMSV}$ on the squeezing amplitude $S\ (dB)$. The deterioration in sensitivity in the practical range of small values of $S$ is noticeable in comparison with those present in Figures 2 and 3. The uncertainty in estimating the phase shift lies in the range from the Heisenberg limit (dashed lines) to SQL (dash-dot lines).

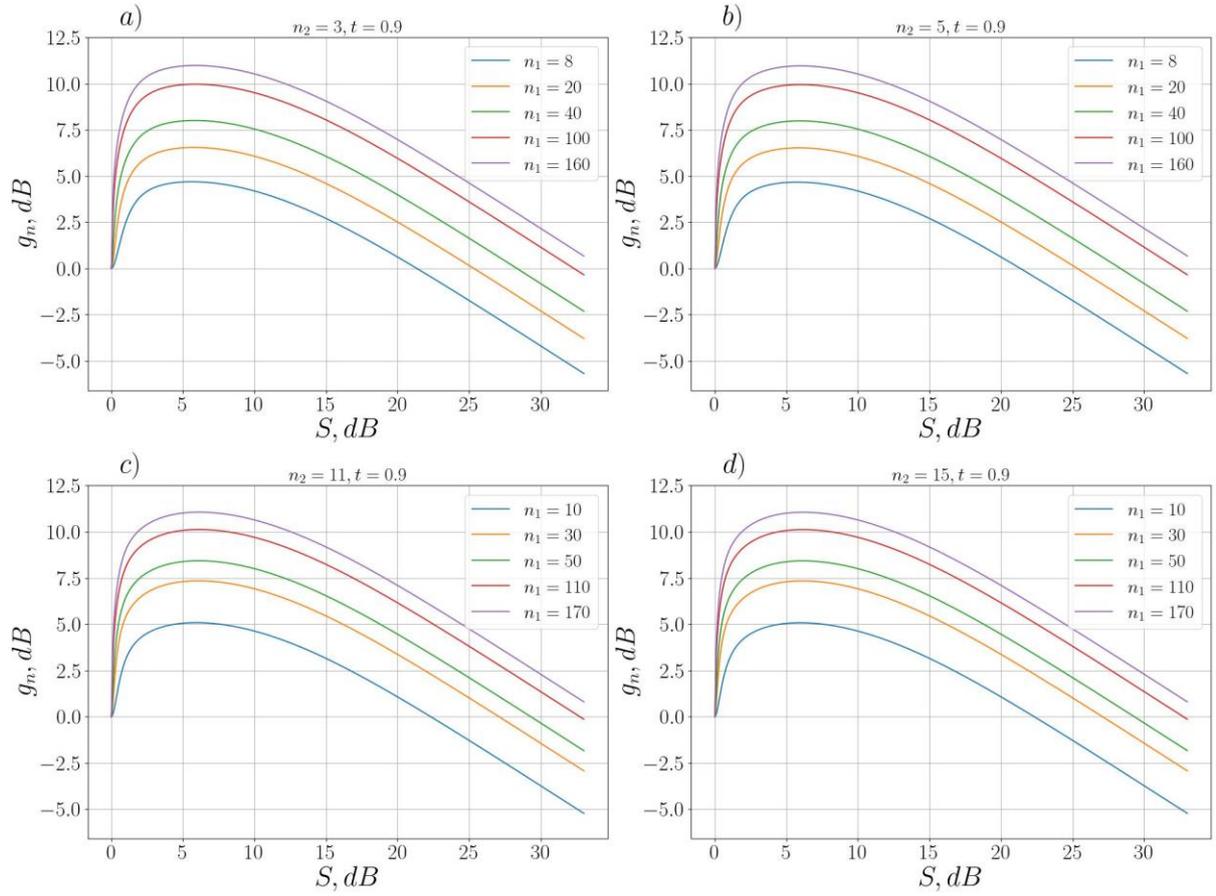

**Fig. 5(a-d).** Gain sensitivity $g_{n_1,n_2/n_1,SMSV}$ ($dB$) (for simplicity, designated as $g_n$ along the vertical axis) as a function of squeezing $S$ ($dB$) of the original SMSV states at different values $n_1$ and $n_2$ and the same transmission amplitude $t$ of the BSs used for construction of states (4) and (14).

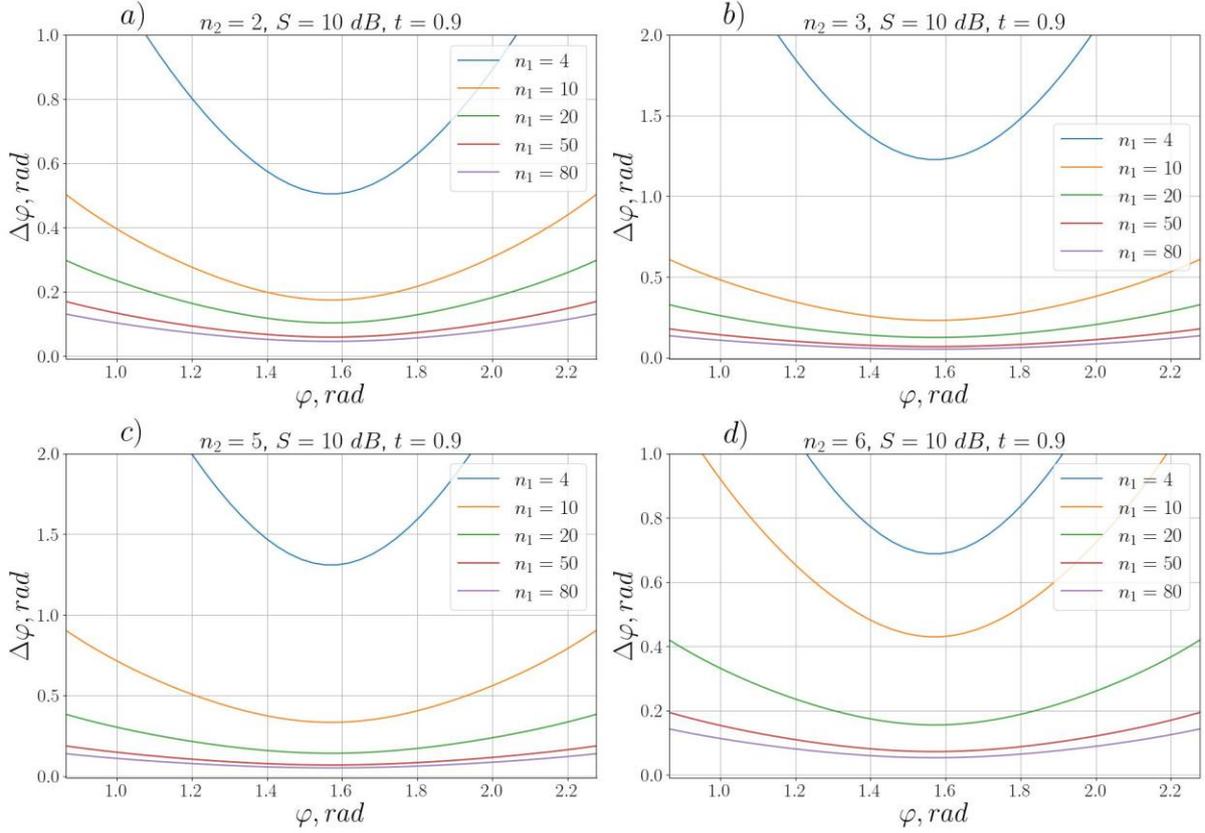

**Fig. 6(a-d).** Sensitivity of phase estimation obtained with intensity difference measurement of the measurement-induced CV states (4) in dependency on the unknown phase shift $\varphi$ for different numbers of the subtracted photons $n_1$ in first and $n_2$ second measurement channels under other identical conditions $S$ and $t$. The minimum phase estimation error is observed at $\varphi = \pi/2$.

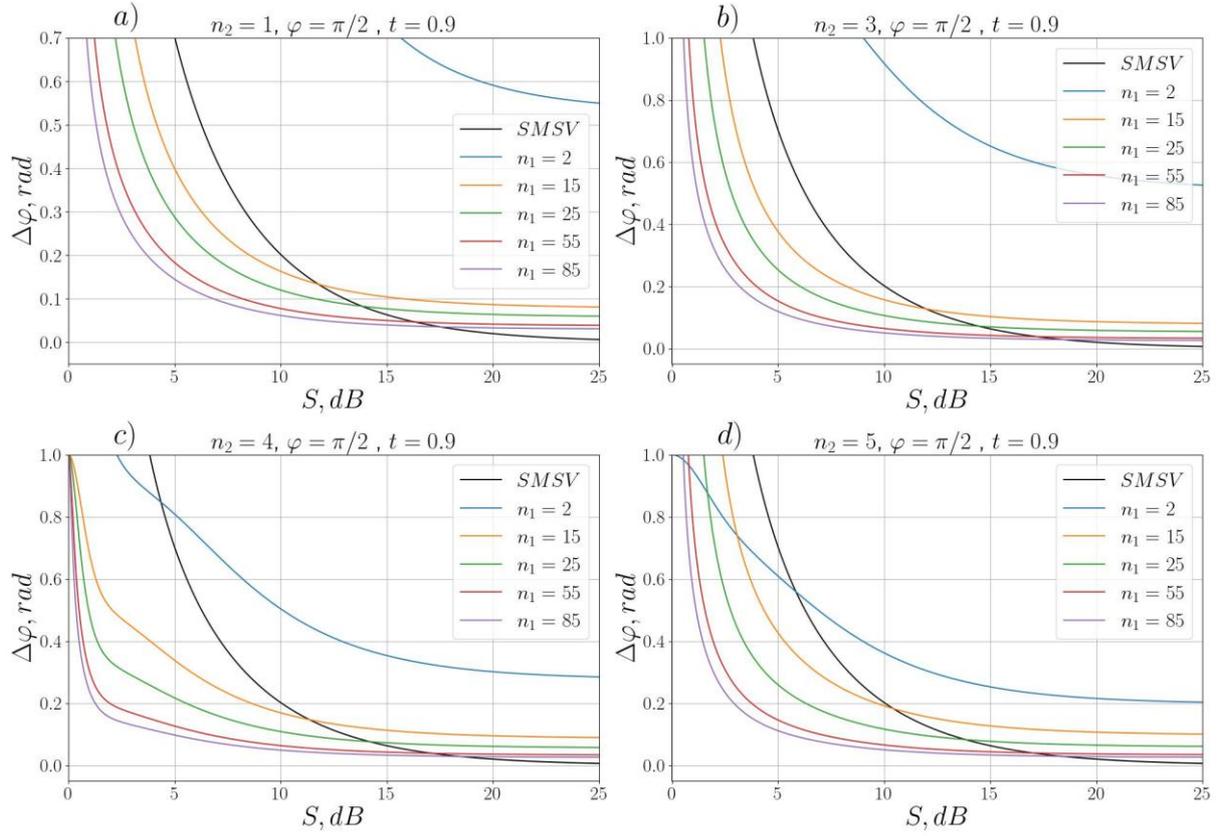

**Fig. 7(a-d).** MZ phase uncertainty $\Delta\varphi$ of phase shift $\varphi = \pi/2$ as a function of squeezing amplitude $S$ ($dB$) with $t$ and $n_2$ being equal for the corresponding set of plots. The dependencies are plotted for different values of $n_1$. QCR bound for the state (6) is also demonstrated on the graphs in black. There are regions of $S$ where a greater sensitivity is achievable when measuring the intensity difference of the measurement-induced CV states (4) on compared even with QCR bound of the initial ones (6) from which they come.